\documentclass{appolb}%

\usepackage{epsfig,amssymb,amsmath}

\begin{document}

%\begin{center}

%\vskip 0.5 cm

%\hfill hep-ph/0504186

%\hfill \hfill UW/PT 04-25

\title{In Search of Lonely Top Quarks at the Tevatron
\thanks{Presented by S.D. Ellis at the XXXIVth International Symposium on 
Multiparticle Dynamics, July 26-August 1, 
2004, Sonoma County, California, USA}}

\author{Matthew T. Bowen, Stephen D. Ellis \\ and Matthew J. Strassler
\address{Department of Physics, P.O Box 351560 \\ 
University of Washington, Seattle, WA 98195  USA}}

\maketitle

\begin{abstract}
Single top-quark production, via weak-interaction %\maketitle
processes, is an important
test of the standard model, potentially sensitive to new physics. However,
this measurement is much more challenging at the Tevatron than originally
expected. We reexamine this process and suggest new methods, using shape
variables, that can supplement the methods that have been discussed
previously. 

\end{abstract}

%\noindent\strut\kern18pt{\small \textrm{PACS numbers: 14.65.Ha, 13.85.Ni,
%13.85.Qk, 13.87.Ce}} \vskip 0.5 cm

%\PACS{14.65.Ha, 13.85.Ni,13.85.Qk, 13.87.Ce}

%\pagestyle{plain}

The electroweak production of single top quarks is an important standard model
process which the Tevatron is guaranteed to be able to study. This reaction,
which has been investigated previously \cite{ssw,OTHERt}, is interesting both
because it provides a direct measurement of the $V_{tb}$ CKM\ element and
because it is sensitive to deviations of top quark physics from standard model
predictions\cite{newsum}. Limits on single top production from 
Run I at the Tevatron have
been published \cite{cdfrunone,d0runone}, and the first Run II limits have
appeared \cite{cdfruntwo,d0runtwo}. This discussion will provide a brief
overview on our recent work on this subject \cite{BES}. 
The executive summary is
that a) this process will be more difficult to study experimentally than 
previously thought, b) we have developed new
analysis methods, which will improve the significance of the
measurement, and c) even with our new analysis methods the
unambiguous detection of this process remains a considerable challenge.
In particular, we suggest making use of more of the information
encoded in the shape of the signal, in a way that will be less sensitive to
systematic errors than a simple counting experiment. On the other hand, we
also show that the size of and the uncertainties in the $W$-plus-jets
background cause serious problems that at present make the measurement
difficult at best.%

%TCIMACRO{\FRAME{ftbphFU}{4.3007in}{1.497in}{0pt}{\Qcb{Single top quark
%production via a) an $s$-channel $W$, b) a $t$-channel $W$.}}{\Qlb{tb}%
%}{tbboth.eps}{\special{ language "Scientific Word";  type "GRAPHIC";
%display "USEDEF";  valid_file "F";  width 4.3007in;  height 1.497in;
%depth 0pt;  original-width 6.3737in;  original-height 5.7294in;
%cropleft "0";  croptop "1";  cropright "1";  cropbottom "0";
%filename 'tbboth.eps';file-properties "XNPEU";}}}%
%BeginExpansion
\begin{figure}
[ptbh]
\begin{center}
\includegraphics[
height=1.497in,
width=4.3007in
]%
{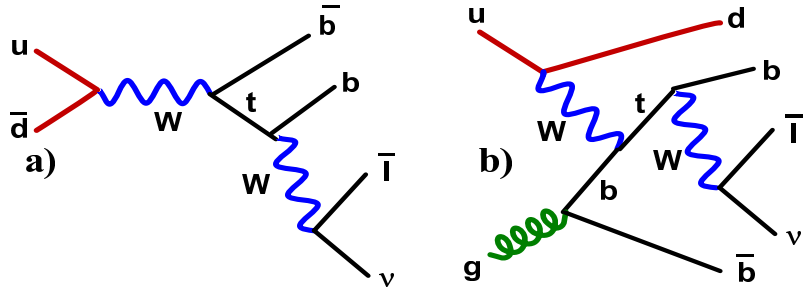}%
\caption{Single top quark production via a) an $s$-channel $W$, b) a
$t$-channel $W$.}%
\label{tb}%
\end{center}
\end{figure}
%EndExpansion

At Fermilab energies, the
``$tb$'' production of a top quark and bottom
antiquark by an $s$-channel $W$ boson, as can occur through the diagram in
Fig.~\ref{tb}a), has a lower cross-section than ``$tbq$'' 
production via a $t$-channel $W$ boson of a $t$,
$\bar{b}$, and an extra quark jet near the beam axis, as occurs through
diagrams such as that in Fig.~\ref{tb}b). 

From Fig.~\ref{tb}b) one can see that a $tbq$ event has a final partonic state
consisting of at least the following: a charged lepton, a neutrino, $b$ and
$\bar{b}$ quarks, and a light quark. Thus, in selecting $t$-channel signal
events, one asks for (a) one $b$-tagged jet, (b) significant missing energy
(from the neutrino), (c) one and only one isolated $e^{\pm}$ or $\mu^{\pm}$,
(d) at least one non-$b$-tagged jet. Typically, the highest-$p_{T}$
non-$b$-tagged jet in a $t$-channel event is that from the light quark. Also,
in typical events a $t$ quark can be reconstructed from the tagged jet and a
$W$, itself reconstructed from the charged lepton and the missing energy. 
The $tb$ process has a $b$ quark jet and
a $\bar{b}$ jet, along with a lepton and a neutrino. In a significant number
of events, one of the two $b$-jets will not be tagged, so that the same
criteria used for $tbq$ will have moderately high efficiency for this
process as well. 

The main backgrounds to single top production, which all can imitate the
signature just described, are \cite{ssw} (1) \textquotedblleft$t\bar{t}%
$\textquotedblright, top quark pair production, primarily from events where
only one of the top quarks decays leptonically; (2) \textquotedblleft
QCD\textquotedblright\ events with fake electrons or muons, or with real muons
from heavy flavor decays; and, most problematic, (3) \textquotedblleft$Wj^{n}%
$\textquotedblright\ events with a leptonically-decaying $W$ plus some number
$n\geq2$ of quark or gluon jets. Processes of all three types occur at rates
well above the signal rate and so we must provide specialized cuts and other
methods to reduce their contribution to the final event sample. For example,
the $t\bar{t}$ background events tend to exhibit considerably more transverse
energy than true single top events, to have more jets, and to be more
spherical. \ Experience with the QCD\ background suggests that it is generally
reduced to a level comparable with the signal \cite{d0runtwo} by fairly
standard cuts and we will not focus on it here, except to note that our event
shape methods will tend to also substantially reduce the QCD component of the
event sample. \ The $W$-plus-jets background is much more complicated. The
$Wj^{n}$ events potentially entering the sample consist of a real $W$ boson
decaying leptonically, and at least two other quarks or gluons in the final
state. While the $Wj^{n}$ events do tend to have smaller energetics than
single top, and, of course, have no reconstructible $t$ quark; the number of
events is so large, and the energy resolution at Fermilab is sufficiently
broad, that the $Wj^{n}$ events form a large and problematic background to
single-top production. Reducing the systematic error on the prediction and/or
measurement of this process is essential for success.

To model both signal and background, we have used MadEvent \cite{madevent} to
generate events, Pythia \cite{pythia} to then simulate showering and
hadronization, and PGS \cite{pgs} to act as a fast detector
simulation. (See \cite{BES} for the details of the calculations.) 
Jets containing a $b$ quark (either
perturbatively or produced during showering) are taken to be tagged with an
efficiency of the form $0.5\tanh\left(  p_{T}/36\ \text{GeV}\right)  $, where
$p_{T}$ is the transverse momentum of the jet. Jets containing a $c$ quark
(either perturbatively or produced during showering) are taken to be tagged
with a rate of the form $0.15\tanh\left(  p_{T}/42\ \text{GeV}\right)  $;
while jets containing no heavy flavor are taken to be mistagged with a rate of
the form $0.01\tanh\left(  p_{T}/80\ \text{GeV}\right) $.

We have chosen for this study to cut also on the quantity
\[
H_{T}=\sum_{\mathrm{jets}}(p_{T})_{i}+(p_{T})_{\ell}+\slash\hskip-.11inE_{T}%
\ ,
\]
which, compared to the signal, is smaller for $Wj^{n}$ and larger for $t\bar{t}$.
In this expression the sum is over all jets with $p_{T}>20$ GeV and $|\eta|<3.5$,
$(p_{T})_{i}$ is the magnitude of the transverse momentum of the $i^{th}$ jet,
$(p_{T})_{\ell}$ that of the lepton, and $\slash\hskip-.11inE_{T}$ is the
missing transverse energy in the event. Since
the signal involves a $t$-quark, we also impose a requirement that the
invariant mass of the lepton, neutrino, and the leading tagged jet be
within a window around the top quark mass.

We find that such cuts cannot decrease the backgrounds
to the point that they are comparable to the signal. Two choices of
\textquotedblleft intermediate\textquotedblright and \textquotedblleft
hard\textquotedblright\ cuts are indicated in Table \ref{cuts2i}. The
resulting numbers of expected events for an integrated luminosity of 3
fb$^{-1}$, summing over $e^{\pm}$ and $\mu^{\pm}$ (and thus including both $t$
and $\overline{t}$), can be seen in Table \ref{results1}. Consistent with
\cite{ssw}, we find that while all of the cuts contribute to the background
reduction, the $Wj^{n}$ channel is reduced primarily by a combination of the
stiffer $p_{T}$ cuts and the \textquotedblleft$m_{t} $\textquotedblright\ cut,
while the $t\overline{t}$ background is affected primarily by the upper
$H_{T}$ cut.  While the intermediate cuts reveal a signal to background ratio 
of 1:7.4, this improves to 1:4.9 using the hard cuts. (The basic cuts used
as a starting point in \cite{BES} yield a signal to background ratio of
approximately 1:21.)
This result is, at best, disappointing and more pessimistic than \cite{ssw}. 
(For a detailed comparison
with the results of \cite{ssw} see \cite{BES}.)
We are concerned that systematic errors in
the understanding of the background will plague a direct counting experiment
at a level that will make any claims of discovery suspect. 

\begin{table}[ptb]
\medskip
\par
\begin{center}%
\begin{tabular}
[c]{|c|c|c||c|c|}\hline\hline
\multicolumn{1}{||c|}{Item} & \multicolumn{1}{||c|}{$p_{T}$} &
\multicolumn{1}{||c||}{$\left\vert \eta\right\vert $} &
\multicolumn{1}{||c|}{$p_{T}$} & \multicolumn{1}{||c||}{$\left\vert
\eta\right\vert $}\\\hline\hline
$\ell^{\pm}$ & $\geq15$ GeV & $\leq2$ & $\geq15$ GeV & $\leq2$\\\hline
MET $\left(  \nu\right)  $ & $\geq15$ GeV & - & $\geq15$ GeV & -\\\hline
jet $\left(  b\text{-tag}\right)  $ & $\geq20$ GeV & $\leq2$ & $\geq60$ GeV &
$\leq2$\\\hline
jet $\left(  \text{no }b\text{-tag}\right)  $ & $\geq20$ GeV & $\leq3.5$ &
$\geq30$ GeV & $\leq3.5$\\\hline\hline
& Min & Max & Min & Max\\\hline
$H_{T}$ & $\geq180$ GeV & $\leq250$ GeV & $\geq180$ GeV & $\leq250$
GeV\\\hline
\textquotedblleft$m_{t}$\textquotedblright & $\geq160$ GeV & $\leq190$ GeV &
$\geq160$ GeV & $\leq190$ GeV\\\hline
\end{tabular}
\end{center}
\caption{Representative \textquotedblleft intermediate\textquotedblright\ cuts
(columns 2 and 3) and \textquotedblleft hard\textquotedblright\ cuts (columns
4 and 5).}%
\label{cuts2i}%
\end{table}

\begin{table}[ptb]
\medskip
\par
\begin{center}%
\begin{tabular}
[c]{|c|c|c|}\hline\hline
\multicolumn{1}{||c|}{Channel} & 
\multicolumn{1}{||c|}{\textquotedblleft Intermediate \textquotedblright\ Cuts} &
\multicolumn{1}{||c||}{\textquotedblleft Hard\textquotedblright\ Cuts}%
\\\hline\hline
$tbq$ & $67$ & $30$\\\hline
$tb$ &  $27$ & $13$\\\hline
$t\bar{t}$ & $140$ & $57$\\\hline
$Wjj$ & $550$ & $152$\\\hline\hline
$\left(  tbq+tb\right)  /\left(  t\bar{t}+Wjj\right)  $ & $0.14$ &
$0.21$\\\hline
\end{tabular}
\end{center}
\caption{Numbers of events for 3 fb$^{-1}$ (summed over $t$ and $\bar{t}$, $e
$ and $\mu$ channels) for the two sets of cuts in Table \ref{cuts2i}.}%
\label{results1}%
\end{table}

Under the assumption that a counting experiment is indeed insufficient, we
turn to observables that (as in \cite{cdfrunone,cdfruntwo}) make use of other
aspects of the signal. For the dominant $tbq$ production process the light quark,
forward jet is a distinctive signature, which the backgrounds do not share.
We make use of this feature in our analysis 
of both signal and background events
by including in our event definition
the properties of ``the jet'', defined to be the highest-$p_{T}$
non-$b$-tagged jet.  
Since the $tbq$
process arises from an initial state light quark or antiquark scattering off a
gluon, the $tbq$ system is typically boosted in the direction of the initial
quark. Moreover the motion of the final state light quark is typically
in the same direction as its parent quark (and the proton). 
(The reverse is true for $\bar{t}$ production.) The structure of
the electroweak interactions in both the production and decay
of the top quark tends to
align the momentum direction of ``the jet'' with that of the 
final-state charged lepton. Thus the momentum vectors of the lepton and ``the
jet'' are correlated as a result of both kinematics and spin
polarization effects.

The $t\bar{t}$ process shares none of these properties. At tree-level,
$t\bar{t}$ is separately C and P invariant, and shows none of the above
asymmetries. The parity-invariance is violated at the next order 
\cite{kuhnstudy}, at a level both small
and calculable. Moreover, there should be no strong correlation between the
momenta of the lepton and the jets in the final state. Similar considerations 
apply, to a good approximation, to those QCD events
which might pass our cuts, mainly events with fake leptons or with isolated
leptons from heavy flavor decays. (See \cite{BES} for additional discussion of
this point.)

The asymmetries and correlations in $Wj^{n}$ are similar to those of $tbq$
(unfortunately) although less pronounced. 
As with a $t$ quark, a $W^{+}$ is most likely to be produced moving in
the proton direction. This leads to parity asymmetries which, though
relatively small, are still quite large in absolute size compared to the
signal. Correlations between the lepton and ``the jet''
are nonzero, though relatively small. Unfortunately, the size of the
asymmetries and correlations in $Wj^{n}$ appears to be very sensitive to
assumptions, cuts, Monte Carlo parameters, and tagging, and will be a source
of significant systematic error.

To make the use of these special properties of the signals and backgrounds we
consider two-dimensional distributions in
pseudo-rapidity. Since the $p\bar{p}$ initial state of the Tevatron is a CP
eigenstate, the differential cross-section $d^{2}\sigma
^{+}/d\eta_{j}d\eta_{\ell}$ describing the rapidity distributions
of ``the jet'' and the  
charged lepton in events with a \textit{positively} charged lepton, 
and the corresponding distribution $d^{2}\sigma
^{-}/d\eta_{j}d\eta_{\ell}$ for processes with a \textit{negatively} charged
lepton, \textit{must} be CP invariant:
\[
{\frac{d^{2}\sigma^{+}}{d\eta_{j}d\eta_{\ell}}}(\eta_{j},\eta_{\ell}%
)={\frac{d^{2}\sigma^{-}}{d\eta_{j}d\eta_{\ell}}}(-\eta_{j},-\eta_{\ell})\ .
\]
Consequently, we can combine data from positively and negatively charged
leptons by defining a \textit{lepton-charge-weighted pseudo-rapidity},
$\hat{\eta}_{j}=Q_{\ell}\eta_{j}$, $\hat{\eta}_{\ell}=Q_{\ell}\eta_{\ell}$,
where $Q_{\ell}$ is the lepton charge. The remainder of this
discussion is based on the explicitly CP-invariant differential cross section
\[
{\frac{d^{2}\sigma}{d\hat{\eta}_{j}d\hat{\eta}_{\ell}}}(\hat{\eta}_{j}%
,\hat{\eta}_{\ell})\equiv{\frac{d^{2}\sigma^{+}}{d\eta_{j}d\eta_{\ell}}}%
(\eta_{j}=\hat{\eta}_{j},\eta_{\ell}=\hat{\eta}_{\ell})+{\frac{d^{2}\sigma
^{-}}{d\eta_{j}d\eta_{\ell}}}(\eta_{j}=-\hat{\eta}_{j},\eta_{\ell}=-\hat{\eta
}_{\ell})\ .
\]

\begin{figure}[ptb]
\begin{center}
\includegraphics[
height=3.78in,
width=4.9in
]{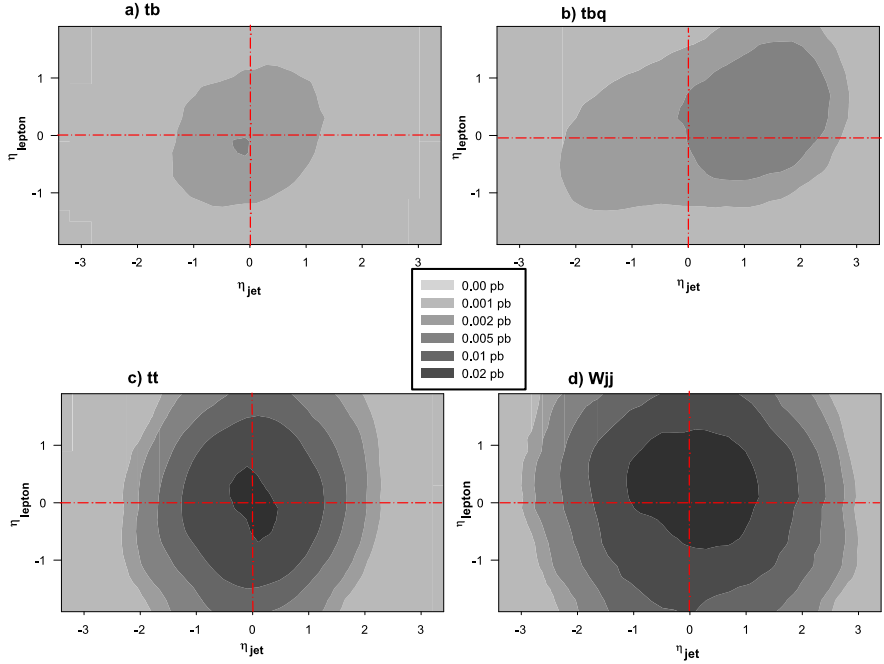}
\end{center}
\caption{Differential cross-section ($d^{2}\sigma/d\hat{\eta}_{j}d\hat{\eta
}_{\ell}$, summed over $t$ and $\overline{t}$, $e$ and $\mu$) for the a) $tb$
channel b) $tbq$ channel, c) $t\overline{t}$ channel, and d) $Wjj$ channel.
}%
\label{all4sig}%
\end{figure}

To illustrate the characteristic properties of the various channels we will
study\cite{BES} a simulated event sample defined by cuts somewhat more 
\textquotedblleft relaxed\textquotedblright\ than those of Table
\ref{cuts2i}, keeping a larger fraction of both the signal and the 
background.
The simulated differences in shape are
summarized in the contour plots of Fig.~\ref{all4sig}, which give the
distributions ($d^{2}\sigma/d\hat{\eta}_{j}d\hat{\eta}_{\ell}$) of the various
processes in the $(\hat{\eta}_{j},\hat{\eta}_{\ell})$ plane.

Figures \ref{all4sig}a and \ref{all4sig}c illustrate the relative
symmetry of the $tb$ and $t\bar{t}$ channel, to be compared to the small
asymmetry in the $Wjj$ channel in Fig.~\ref{all4sig}d and the large asymmetry
in the $tbq$ signal channel of Fig.~\ref{all4sig}b. In particular, the $tbq$
signal lies dominantly in the B quadrant, where the four quadrants of the
$(\hat{\eta}_{j},\hat{\eta}_{\ell})$ plane are labeled in the form
$\frac{A\ \,|\ B}{C\,\,\ |\ D}$. \ Meanwhile $tb$ is smaller with a less
distinctive shape. By contrast, $t\bar{t}$ is large though symmetrically
distributed, and $Wj^{n}$ is large but has small asymmetries between the four
quadrants, with slightly more cross-section in the A and B quadrants than in C
and D.

To capture quantitatively these differences in shape between signal and
background, we define three orthogonal functions in the $(\hat{\eta}_{j}%
,\hat{\eta}_{\ell})$ plane. Any differential cross-section
can be written as a sum of three components
\begin{equation}
{\frac{d^{2}\sigma_{{}}}{d\hat{\eta}_{j}d\hat{\eta}_{\ell}}}(\hat{\eta}%
_{j},\hat{\eta}_{\ell})=\bar{F}(\hat{\eta}_{j},\hat{\eta}_{\ell})+F_{+}%
(\hat{\eta}_{j},\hat{\eta}_{\ell})+F_{-}(\hat{\eta}_{j},\hat{\eta}_{\ell}),
\label{sigmaFFF}%
\end{equation}
where the components are of the form%
\begin{align}
\bar{F}(\hat{\eta}_{j},\hat{\eta}_{\ell})  &  \equiv{\frac{1}{4}}\left[
{\frac{d^{2}\sigma_{{}}}{d\hat{\eta}_{j}d\hat{\eta}_{\ell}}}(\hat{\eta}%
_{j},\hat{\eta}_{\ell})+{\frac{d^{2}\sigma_{{}}}{d\hat{\eta}_{j}d\hat{\eta
}_{\ell}}}(-\hat{\eta}_{j},-\hat{\eta}_{\ell})\right. \nonumber \\
&  \left.  +{\frac{d^{2}\sigma_{{}}}{d\hat{\eta}_{j}d\hat{\eta}_{\ell}}}%
(\hat{\eta}_{j},-\hat{\eta}_{\ell})+{\frac{d^{2}\sigma_{{}}}{d\hat{\eta}%
_{j}d\hat{\eta}_{\ell}}}(-\hat{\eta}_{j},\hat{\eta}_{\ell})\right],\label{Fbardef}
\end{align}%
\begin{align}
F_{+}(\hat{\eta}_{j},\hat{\eta}_{\ell})  &  \equiv{\frac{1}{4}}\left[
{\frac{d^{2}\sigma_{{}}}{d\hat{\eta}_{j}d\hat{\eta}_{\ell}}}(\hat{\eta}%
_{j},\hat{\eta}_{\ell})+{\frac{d^{2}\sigma_{{}}}{d\hat{\eta}_{j}d\hat{\eta
}_{\ell}}}(-\hat{\eta}_{j},-\hat{\eta}_{\ell})\right. \nonumber \\
&  \left.  -{\frac{d^{2}\sigma_{{}}}{d\hat{\eta}_{j}d\hat{\eta}_{\ell}}}%
(\hat{\eta}_{j},-\hat{\eta}_{\ell})-{\frac{d^{2}\sigma_{{}}}{d\hat{\eta}%
_{j}d\hat{\eta}_{\ell}}}(-\hat{\eta}_{j},\hat{\eta}_{\ell})\right],\label{Fplusdef}
\end{align}%
\begin{equation}
F_{-}(\hat{\eta}_{j},\hat{\eta}_{\ell})\equiv\frac{1}{2}\left[  {\frac
{d^{2}\sigma_{{}}}{d\hat{\eta}_{j}d\hat{\eta}_{\ell}}}(\hat{\eta}_{j}%
,\hat{\eta}_{\ell})-{\frac{d^{2}\sigma_{{}}}{d\hat{\eta}_{j}d\hat{\eta}_{\ell
}}}(-\hat{\eta}_{j},-\hat{\eta}_{\ell})\right].  \label{Fminusdef}%
\end{equation}
These functions are orthogonal in the sense that the
integral of any (nonidentical) pair over any symmetrically-defined region of
the $(\hat{\eta}_{j},\hat{\eta_{\ell}})$ plane vanishes. Most importantly, the
functions provide important physical information about the symmetry properties
of the differential cross-section. $\bar{F}$ and $F_{+}$ are parity-even while
$F_{-}$ is parity-odd; thus $F_{-}=0$ (within statistics) for any parity-even
distribution. Meanwhile $F_{+}$ will also
vanish if the distribution is parity-even \textit{and} the leptons and jets
are uncorrelated. By construction
$\bar{F}$ and $F_{+}$ have four-way symmetry under reflection 
in the $(\hat{\eta}_{j},\hat{\eta}_{\ell})$ plane, while $F_{-}$ has two-way 
symmetry (quadrants A
and D are related, as are B and C, but quadrants A and B are independent).

The fact that the signal has strong asymmetries and
correlations not present in the backgrounds yields the following expectations,
\begin{align*}
\bar{F}^{t\bar{t}}  &  \gg F_{\pm}^{t\bar{t}}\ ;\ \bar{F}^{Wj^{n}}>F_{\pm
}^{Wj^{n}}\ ;\ \bar{F}^{tbq}\sim F_{\pm}^{tbq}\ ,\\
\bar{F}^{t\bar{t}}  &  \sim\bar{F}^{Wj^{n}}\gg\bar{F}^{tbq}\ ,\\
F_{+}^{t\bar{t}}  &  \sim F_{+}^{Wj^{n}}\sim F_{+}^{tbq}\ ,\ F_{-}^{t\bar{t}%
}\ll F_{-}^{Wj^{n}}\sim F_{-}^{tbq}\ .
\end{align*}
Thus the backgrounds are very large only in $\bar{F}$, while the signal has a
much larger role to play in the other functions, especially away from the
center of the $(\hat{\eta}_{j},\hat{\eta}_{\ell})$ plane. 

\begin{figure}[ptbh]
\begin{center}
\includegraphics[
height=3.78in,
width=4.9in
]{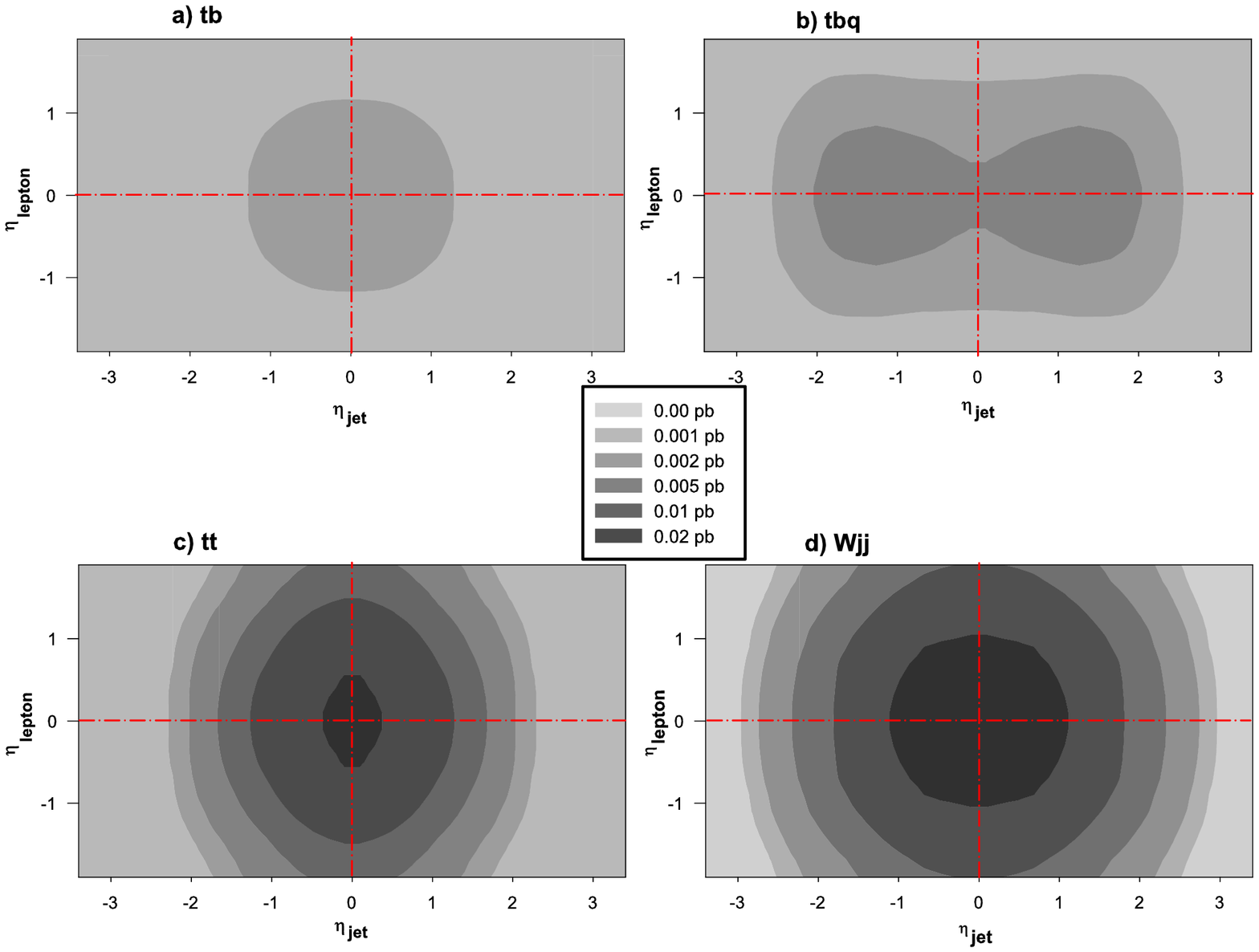}
\end{center}
\caption{Contour plots for the function $\overline{F}\left(  \hat{\eta}%
_{j},\hat{\eta}_{\ell}\right)  $ for a) $tb$, b) $tbq$, c) $t\overline{t}$,
and d) $Wjj$ channels (summed over $t$ and $\overline{t}$, $e$ and $\mu$). }%
\label{allfbar}%
\end{figure}

\begin{figure}[ptb]
\begin{center}
\includegraphics[
trim=0.000000in 0.000000in -0.106586in -0.132145in,
height=3.811in,
width=4.9in
]{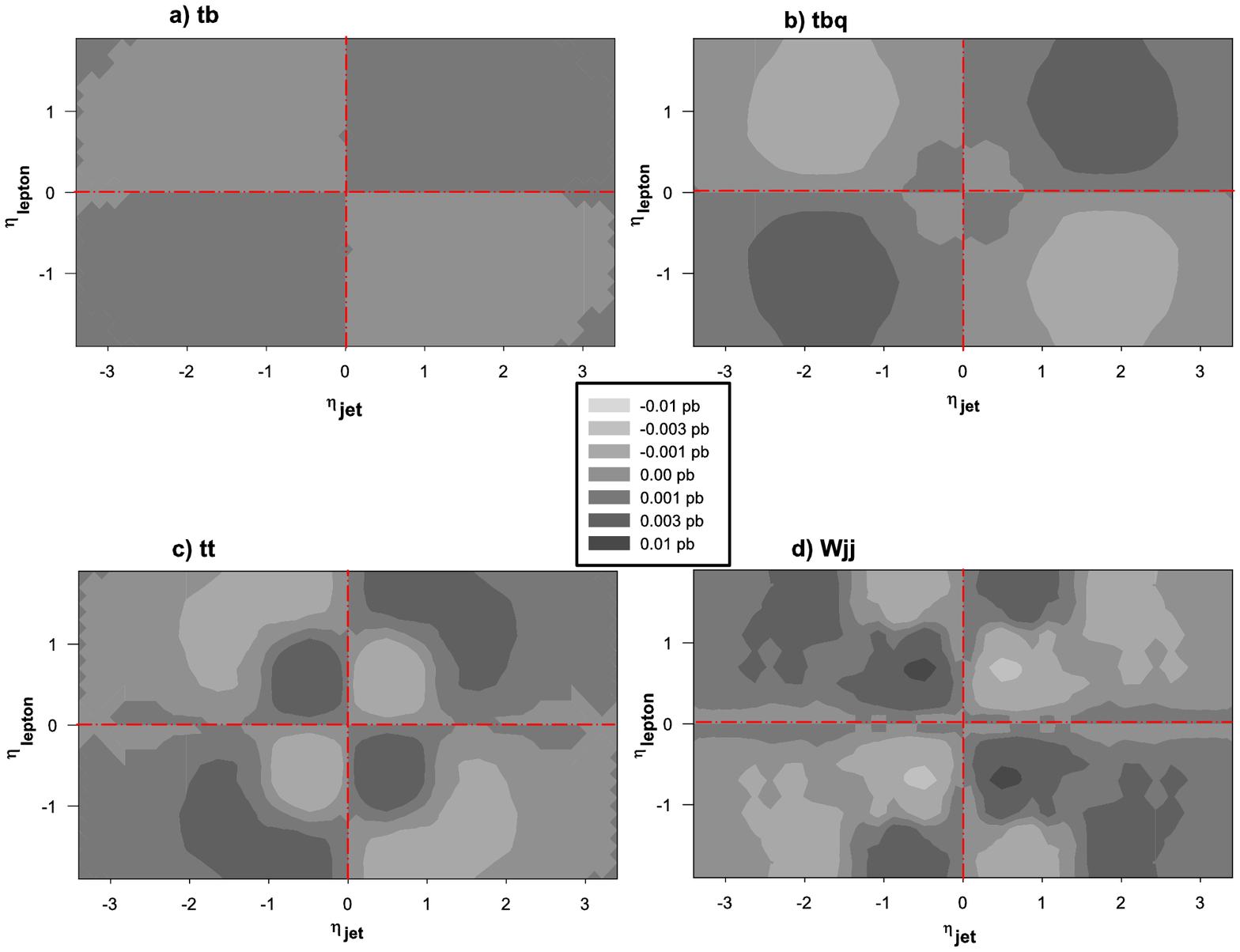}
\end{center}
\caption{Contour plots for the function $F_{+}\left(  \hat{\eta}_{j},\hat
{\eta}_{\ell}\right)  $ for a) $tb$, b) $tbq$, c) $t\overline{t}$, and d)
$Wjj$ channels (summed over $t$ and $\overline{t}$, $e$ and $\mu$). }%
\label{allfsym}%
\end{figure}

\begin{figure}[ptbh]
\begin{center}
\includegraphics[
height=3.78in,
width=4.9in
]{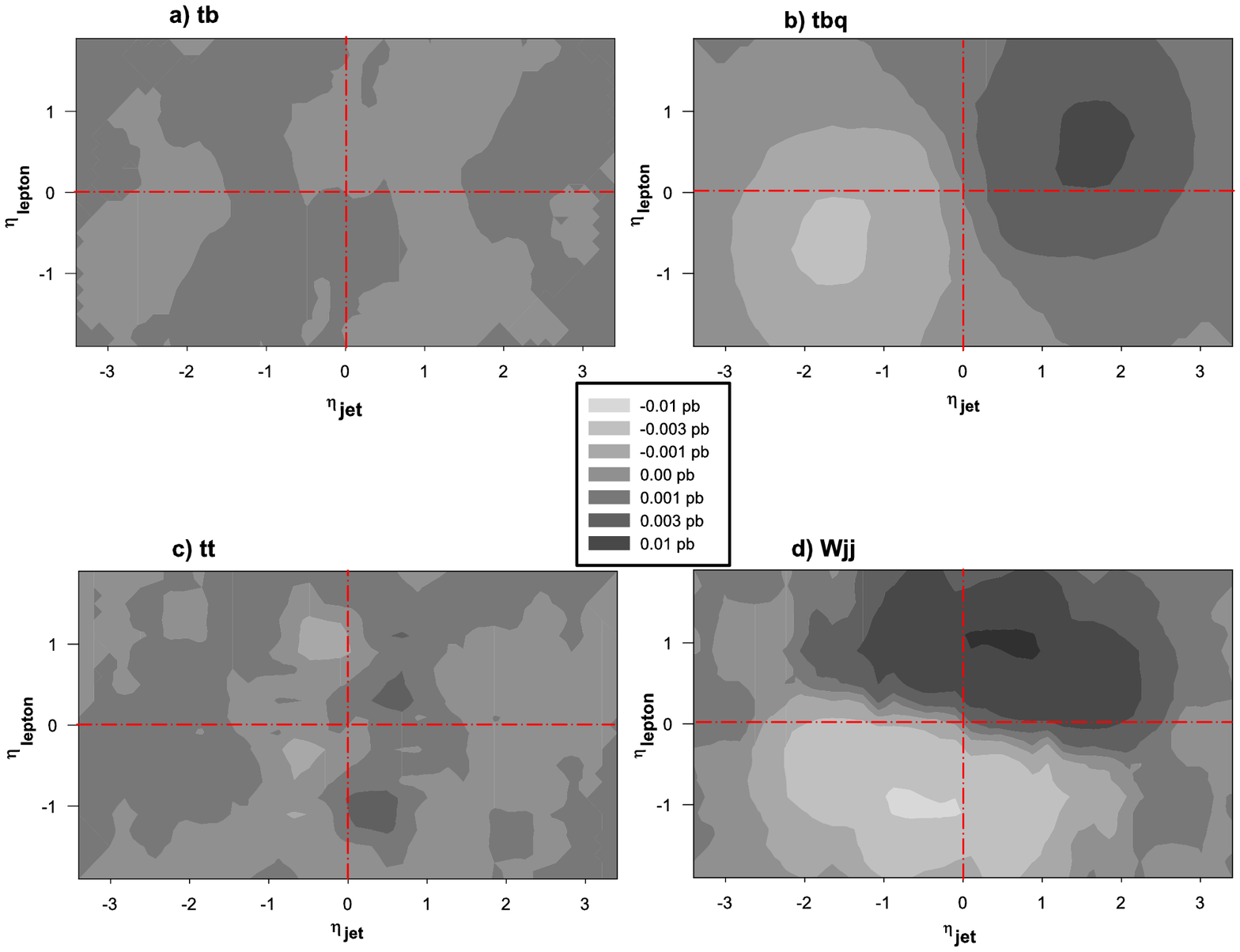}
\end{center}
\caption{Contour plots for the function $F_{-}\left(  \hat{\eta}_{j},\hat
{\eta}_{\ell}\right)  $ for a) $tb$, b) $tbq$, c) $t\overline{t}$, and d)
$Wjj$ channels (summed over $t$ and $\overline{t}$, $e$ and $\mu$). }%
\label{allfasym}%
\end{figure}

These expectations are illustrated in Figs.~\ref{allfbar}, \ref{allfsym} and
\ref{allfasym}, where the functions $\bar{F},F_{+},F_{-}$ are shown, for $tb$,
$tbq$, $t\bar{t}$, and $Wj^{n}$. Note, in particular, the presence 
of the symmetry properties discussed above.

Thus we are led (see \cite{BES} for more details) to 
suggest the following general approach.
First construct, for both the data and the Monte Carlo simulations,
the $\bar{F}$, $F_{+}$ and $F_{-}$ functions. Using these functions,
as well as information obtained from other measurements, we systematically
test our understanding of each process. \ The $\bar{F}$ function allows a
measurement of the sum of the backgrounds without much contamination from
signal.  Assuming that the separation of $Wj^{n}$ from $t\bar{t}$ can be
obtained using the fact that the $t\bar{t}$ process can be measured and
predicted with reasonable accuracy, we
can then cross-check our understanding of the $Wj^{n}$
background using part of the $F_{-}$ distribution (located roughly in quadrant
A) where the signal is negligible. Finally, one can attempt to measure the
signal from $F_{+}$ and from another part (located largely in quadrant B) of
the $F_{-}$ distribution.

A complete analysis of the above ideas has been carried out in \cite{BES}, including
estimates of the associated uncertainties, both statistical and
systematic. 
The primary lesson is that, even when using the shape variables discussed above,
the contributions of the backgrounds, especially
$Wj^{n}$, cannot be rendered truly small compared to the signal. Thus it is
essential that we understand the backgrounds with high precision, including
their shapes as employed here. This last point is especially challenging for
the $Wj^{n}$ case, which receives comparable contributions from
events with tags of $b$ quark jets, tags of $c$ quark jets and mistags
of jets with no heavy flavor. 
Likewise there are comparable contributions
from when the heavy-flavor quark is present in the short-distance scattering
and when it is produced in the subsequent showering.
This means that our ability to accurately simulate the  
background depends on our understanding of a multitude of different 
specific channels, each with their own shapes and their own dependence on cuts, 
parton distributions and other parameters.

In summary, a counting experiment for discovery of electroweak single-top
production appears very challenging. We
have explored the possibility of using the distinctive shape of this process
to separate it from background.  We
have noted that the distributions for $t\bar{t}$ and for QCD 
in the $(\hat{\eta}_{j},\hat{\eta}_{\ell})$ plane are largely
symmetric, while that of the signal is not; the $Wj^{n}$ ($W$-plus-jets)
background is intermediate between them. Constructing the functions $\bar
{F},F_{\pm}$, which have
various symmetry properties, we find that the statistical and systematic errors in the
functions $F_{\pm}$, which are orthogonal to the function $\bar{F}$ 
(that would
be used in a counting experiment), can be brought close to reasonable size
without using extreme cuts. Our method largely removes $t\bar{t}$ and 
QCD events from the observables,
making extreme cuts on $t\bar{t}$ unnecessary, and focusing attention on
$Wj^{n}$ as the main background. The most challenging problem 
is understanding the shape of the $Wj^{n}$ background. 
We believe this calls for a
dedicated study of the the rates, shapes, and flavor content (especially
of bottom versus charm) of both $Wj^{n}$ and $Zj^{n}$, with zero, one and two
tagged jets, blended with theoretically precise predictions, and careful tuning and
cross-checking of Monte Carlo simulations. 

\section*{Acknowledgments}

We thank our colleagues G. Watts, A. Garcia-Bellido, T. Gadfort, A. Haas, H.
Lubatti, and T. Burnett for many useful conversations and direct assistance.
This work was
supported by U.S. Department of Energy grants DE-FG02-96ER40956 and
DOE-FG02-95ER40893, and by an award from the Alfred P. Sloan Foundation.


\bigskip

\begin{thebibliography}{99}                                                                                               %


\bibitem {ssw}T. Stelzer, Z. Sullivan, and S. Willenbrock, Phys.\ Rev.\ D
\textbf{58}, 094021 (1998).

\bibitem {OTHERt}For other previous studies of single top quark production see
S.~S.~Willenbrock and D.~A.~Dicus,
%``Production Of Heavy Quarks From W Gluon Fusion,''
Phys.\ Rev.\ D \textbf{34}, 155 (1986);
%%CITATION = PHRVA,D34,155;%%;
C.~P.~Yuan,
%``A New Method To Detect A Heavy Top Quark At The Tevatron,''
Phys.\ Rev.\ D \textbf{41}, 42 (1990);
%%CITATION = PHRVA,D41,42;%%
S.~Cortese and R.~Petronzio,
%``The Single top production channel at Tevatron energies,''
Phys.\ Lett.\ B \textbf{253}, 494 (1991);
%%CITATION = PHLTA,B253,494;%%
R.~K.~Ellis and S.~J.~Parke,
%``Top quark production by W gluon fusion,''
Phys.\ Rev.\ D \textbf{46}, 3785 (1992);
%%CITATION = PHRVA,D46,3785;%%
D.~O.~Carlson and C.~P.~Yuan,
%``Studying the top quark via the W - gluon fusion process,''
Phys.\ Lett.\ B \textbf{306}, 386 (1993);
%%CITATION = PHLTA,B306,386;%%
T.~Stelzer, S.~Willenbrock,
%``Single-top-quark production via q qbar -> t bbar,''
Phys.\ Lett.\ B \textbf{357}, 125 (1995) [arXiv:hep-ph/9505433]; ``Future
Electroweak Physics as the Fermilab Tevatron: Report of the tev$\_$2000 Study
Group,'' edited by D.~Amidei and R.~Brock, Report No. FERMILAB-Pub-96/082
(1996); A.~P.~Heinson, A.~S.~Belyaev and E.~E.~Boos,
%``Single top quarks at the Fermilab Tevatron,''
Phys.\ Rev.\ D \textbf{56}, 3114 (1997) [arXiv:hep-ph/9612424].
%%CITATION = HEP-PH 9612424;%%


\bibitem {newsum}T. Tait and C.--P. Yuan, Phys.Rev.\ D \textbf{63}, 014018
(2001), hep-ph/0007298.

\bibitem {cdfrunone}CDF Collaboration (D. Acosta et al.), Phys.\ Rev.\ D
\textbf{69}, 052003 (2004); \textit{ibid}. D \textbf{65}, 091102 (2002), hep-ex/0110067.

\bibitem {d0runone}D0 Collaboration (B. Abbot et al.), Phys.\ Rev.\ D
\textbf{63}, 031101 (2001), hep-ex/0008024; D0 Collaboration (V.M. Abazov et
al.), Phys.\ Lett.\ B \textbf{517}, 282 (2001), hep-ex/0106059.

\bibitem {cdfruntwo}CDF Collaboration, (D. Acosta et al.), hep-ex/0410058
(submitted to Phys. Rev. Lett.).

\bibitem {d0runtwo}See ``Search~for~single~top production'' at http://www-d0.fnal.gov/Run2Physics/WWW/results/prelim/TOP/T09/T09.pdf.

\bibitem {BES}M. T. Bowen, S.D. Ellis and M.J. Strassler, \textquotedblleft In
Search of the Lonely Top Quark at the Tevatron,\textquotedblright\ preprint
UW/PT 04-12.


\bibitem {madevent}F.~Maltoni and T.~Stelzer, ``MadEvent: Automatic event
generation with MadGraph,'' JHEP \textbf{0302}, 027 (2003)
[arXiv:hep-ph/0208156].
%%CITATION = HEP-PH 0208156;%%


\bibitem {pythia}T.~Sjostrand, P.~Eden, C.~Friberg, L.~Lonnblad, G.~Miu,
S.~Mrenna and E.~Norrbin,
%``High-energy-physics event generation with PYTHIA 6.1,''
Comput.\ Phys.\ Commun.\ \textbf{135}, 238 (2001) [arXiv:hep-ph/0010017].
%%CITATION = HEP-PH 0010017;%%


\bibitem {pgs}J.S. Conway \textit{et al.} in Proceedings of the Workshop on
Physics at Run II-Supersymmetry/Higgs, Fermilab, 1998, p.39
[arXiv:hep-ph/0010338].
%%CITATION = HEP-PH 0010338;%%



\bibitem {kuhnstudy}J.~H.~Kuhn and G.~Rodrigo,
%``Charge asymmetry of heavy quarks at hadron colliders,''
Phys.\ Rev.\ D \textbf{59}, 054017 (1999) [arXiv:hep-ph/9807420].
%%CITATION = HEP-PH 9807420;%%
See also M. Fischer, \textit{et al.}, Phys.\ Rev.\ D \textbf{65}, 054036
(2002), hep-ph/0101322.





\end{thebibliography}
\end{document}